\pdfoutput=1
\documentclass[12pt,a4paper]{article}

\usepackage{ifthen} 
\newboolean{pdflatex}
\setboolean{pdflatex}{true} 

\newboolean{articletitles}
\setboolean{articletitles}{true} 

\newboolean{uprightparticles}
\setboolean{uprightparticles}{false} 

\usepackage{xspace}
\usepackage{comment}
\usepackage{feynmp-auto}
\usepackage{tikz}
\usepackage{tikz-cd}
\usepackage{subfig}
\usepackage{xfrac} 
\usepackage{bm} 
\usepackage{multirow} 

\newcommand{\LbGDec}{\mbox{\ensuremath{\Lb \to \Lz \g}}\xspace}
\newcommand{\LbJDec}{\mbox{\ensuremath{\Lb \to \Lz \jpsi}}\xspace}

\newcommand{\LDec}{\mbox{\ensuremath{\Lz \to \proton \pim}}\xspace}

\newcommand{\XbGDec}{\mbox{\ensuremath{\Xibm \to \Xim \g}}\xspace}
\newcommand{\XbJDec}{\mbox{\ensuremath{\Xibm \to \Xim \jpsi}}\xspace}
\newcommand{\XDec}{\mbox{\ensuremath{\Xim \to \Lz \pim}}\xspace}

\newcommand{\CFigSizeI}{0.8\textwidth} 
 
\def\paperauthors{LHCb collaboration} 
\def\paperasciititle{Search for the radiative Xib2XiGamma decay} 
\def\papertitle{Search for the \\ radiative \XbGDec decay} 
\def\paperkeywords{{High Energy Physics}, {LHCb}, {b-baryons}, {radiative}} 
\def\papercopyright{\the\year\ CERN for the benefit of the LHCb collaboration} 
\def\paperlicence{CC BY 4.0 licence}
\def\paperlicenceurl{https://creativecommons.org/licenses/by/4.0/}

\usepackage[top=1in, bottom=1.25in, left=1in, right=1in]{geometry}

\usepackage{microtype}
\usepackage{lineno}  
\usepackage{xspace} 
\usepackage{caption} 

\usepackage{graphicx}  
\usepackage{color}
\usepackage{colortbl}
\graphicspath{{./figs/}} 

\usepackage{amsmath} 
\usepackage{amssymb}
\usepackage{amsfonts}
\usepackage{upgreek} 

\newcommand*\patchAmsMathEnvironmentForLineno[1]{%
\expandafter\let\csname old#1\expandafter\endcsname\csname #1\endcsname
\expandafter\let\csname oldend#1\expandafter\endcsname\csname
end#1\endcsname
 \renewenvironment{#1}%
   {\linenomath\csname old#1\endcsname}%
   {\csname oldend#1\endcsname\endlinenomath}%
}
\newcommand*\patchBothAmsMathEnvironmentsForLineno[1]{%
  \patchAmsMathEnvironmentForLineno{#1}%
  \patchAmsMathEnvironmentForLineno{#1*}%
}
\AtBeginDocument{%
\patchBothAmsMathEnvironmentsForLineno{equation}%
\patchBothAmsMathEnvironmentsForLineno{align}%
\patchBothAmsMathEnvironmentsForLineno{flalign}%
\patchBothAmsMathEnvironmentsForLineno{alignat}%
\patchBothAmsMathEnvironmentsForLineno{gather}%
\patchBothAmsMathEnvironmentsForLineno{multline}%
\patchBothAmsMathEnvironmentsForLineno{eqnarray}%
}

\usepackage{hyperxmp}

\usepackage[pdftex,
            pdfauthor={\paperauthors},
            pdftitle={\paperasciititle},
            pdfkeywords={\paperkeywords},
            pdfcopyright={Copyright (C) \papercopyright},
            pdflicenseurl={\paperlicenceurl}]{hyperref}

\usepackage[colorinlistoftodos,textsize=scriptsize]{todonotes}

\usepackage[bottom,flushmargin,hang,multiple]{footmisc}

\usepackage[all]{hypcap} 

\usepackage{xspace} 
\usepackage{upgreek}

\def\lhcb   {\mbox{LHCb}\xspace}

\def\lhc    {\mbox{LHC}\xspace}

\def\MagUp {\mbox{\em Mag\kern -0.05em Up}\xspace}

\ifthenelse{\boolean{uprightparticles}}%
{
 
 \def\Pgamma      {\ensuremath{\upgamma}\xspace}

 \def\Pmu         {\ensuremath{\upmu}\xspace}

 \def\Ppi         {\ensuremath{\uppi}\xspace}

 \def\Ppsi        {\ensuremath{\uppsi}\xspace}

 \def\PDelta      {\ensuremath{\Delta}\xspace}                 
 \def\PXi         {\ensuremath{\Xi}\xspace}                 
 \def\PLambda     {\ensuremath{\Lambda}\xspace}                 
 \def\PSigma      {\ensuremath{\Sigma}\xspace}                 
 \def\POmega      {\ensuremath{\Omega}\xspace}                 
 \def\PUpsilon    {\ensuremath{\Upsilon}\xspace}

 \def\PB      {\ensuremath{\mathrm{B}}\xspace}                 
                  
 \def\PD      {\ensuremath{\mathrm{D}}\xspace}

 \def\PJ      {\ensuremath{\mathrm{J}}\xspace}                 
 \def\PK      {\ensuremath{\mathrm{K}}\xspace}

 \def\PW      {\ensuremath{\mathrm{W}}\xspace}

 \def\Pb      {\ensuremath{\mathrm{b}}\xspace}                 
 \def\Pc      {\ensuremath{\mathrm{c}}\xspace}

 \def\Pi      {\ensuremath{\mathrm{i}}\xspace}

 \def\Pp      {\ensuremath{\mathrm{p}}\xspace}

 \def\Ps      {\ensuremath{\mathrm{s}}\xspace}

 \def\thebaroffset{0.0em}
}
{
 
 \def\Pgamma      {\ensuremath{\gamma}\xspace}

 \def\Pmu         {\ensuremath{\mu}\xspace}

 \def\Ppi         {\ensuremath{\pi}\xspace}

 \def\Ppsi        {\ensuremath{\psi}\xspace}                 
                  
 \mathchardef\PDelta="7101
 \mathchardef\PXi="7104
 \mathchardef\PLambda="7103
 \mathchardef\PSigma="7106
 \mathchardef\POmega="710A
 \mathchardef\PUpsilon="7107
                  
 \def\PB      {\ensuremath{B}\xspace}                 
                  
 \def\PD      {\ensuremath{D}\xspace}

 \def\PJ      {\ensuremath{J}\xspace}                 
 \def\PK      {\ensuremath{K}\xspace}

 \def\PW      {\ensuremath{W}\xspace}

 \def\Pb      {\ensuremath{b}\xspace}                 
 \def\Pc      {\ensuremath{c}\xspace}

 \def\Pi      {\ensuremath{i}\xspace}

 \def\Pp      {\ensuremath{p}\xspace}

 \def\Ps      {\ensuremath{s}\xspace}

 \def\thebaroffset{0.18em}
}
\newcommand{\offsetoverline}[2][\thebaroffset]{\kern #1\overline{\kern -#1 #2}}%

\makeatletter
\ifcase \@ptsize \relax
  \newcommand{\miniscule}{\@setfontsize\miniscule{4}{5}}
\or
  \newcommand{\miniscule}{\@setfontsize\miniscule{5}{6}}
\or
  \newcommand{\miniscule}{\@setfontsize\miniscule{5}{6}}
\fi
\makeatother

\DeclareRobustCommand{\optbar}[1]{\shortstack{{\miniscule (\rule[.5ex]{1.25em}{.18mm})}
  \\ [-.7ex] $#1$}}

\def\mup        {{\ensuremath{\Pmu^+}}\xspace}
\def\mun        {{\ensuremath{\Pmu^-}}\xspace} 

\def\g      {{\ensuremath{\Pgamma}}\xspace}

\def\Wm     {{\ensuremath{\PW^-}}\xspace}

\def\squark    {{\ensuremath{\Ps}}\xspace}

\def\cquark    {{\ensuremath{\Pc}}\xspace}

\def\bquark    {{\ensuremath{\Pb}}\xspace}

\def\pion   {{\ensuremath{\Ppi}}\xspace}
\def\piz    {{\ensuremath{\pion^0}}\xspace}
\def\pip    {{\ensuremath{\pion^+}}\xspace}
\def\pim    {{\ensuremath{\pion^-}}\xspace}

\def\kaon    {{\ensuremath{\PK}}\xspace}

\def\KorKbar {\kern \thebaroffset\optbar{\kern -\thebaroffset \PK}{}\xspace}

\def\Km      {{\ensuremath{\kaon^-}}\xspace}

\def\Kstarz  {{\ensuremath{\kaon^{*0}}}\xspace}

\def\Kstar   {{\ensuremath{\kaon^*}}\xspace}

\def\D       {{\ensuremath{\PD}}\xspace}

\def\DorDbar {\kern \thebaroffset\optbar{\kern -\thebaroffset \PD}\xspace}
\def\Dz      {{\ensuremath{\D^0}}\xspace}

\def\Dp      {{\ensuremath{\D^+}}\xspace}
\def\Dm      {{\ensuremath{\D^-}}\xspace}

\def\DpDm    {\ensuremath{\Dp {\kern -0.16em \Dm}}\xspace}

\def\B       {{\ensuremath{\PB}}\xspace}

\def\BorBbar {\kern \thebaroffset\optbar{\kern -\thebaroffset \PB}\xspace}
\def\Bz      {{\ensuremath{\B^0}}\xspace}

\def\Bd      {{\ensuremath{\B^0}}\xspace}

\def\BdorBdbar {\kern \thebaroffset\optbar{\kern -\thebaroffset \Bd}\xspace}

\def\Bs      {{\ensuremath{\B^0_\squark}}\xspace}

\def\BsorBsbar {\kern \thebaroffset\optbar{\kern -\thebaroffset \Bs}\xspace}

\def\jpsi     {{\ensuremath{{\PJ\mskip -3mu/\mskip -2mu\Ppsi}}}\xspace}

\def\Y#1S{\ensuremath{\PUpsilon{(#1S)}}\xspace}

\def\proton      {{\ensuremath{\Pp}}\xspace}

\def\Lz          {{\ensuremath{\PLambda}}\xspace}

\def\LorLbar     {\kern \thebaroffset\optbar{\kern -\thebaroffset \PLambda}\xspace}

\def\Xires       {{\ensuremath{\PXi}}\xspace}

\def\Xim         {{\ensuremath{\Xires^-}}\xspace}

\def\Omegares    {{\ensuremath{\POmega}}\xspace}

\def\Omegam      {{\ensuremath{\Omegares^-}}\xspace}

\def\Lb           {{\ensuremath{\Lz^0_\bquark}}\xspace}

\def\Xib          {{\ensuremath{\Xires_\bquark}}\xspace}

\def\Xibm         {{\ensuremath{\Xires^-_\bquark}}\xspace}

\def\Omegab       {{\ensuremath{\Omegares^-_\bquark}}\xspace}

\def\BF         {{\ensuremath{\mathcal{B}}}\xspace}
\def\BR         {\BF}

\newcommand{\decay}[2]{\ensuremath{#1\!\to #2}\xspace} 

\def\to                 {\ensuremath{\rightarrow}\xspace}

\def\order   {{\ensuremath{\mathcal{O}}}\xspace}

\def\qsq       {{\ensuremath{q^2}}\xspace}

\def\BdKstGam     {\decay{\Bd}{\Kstarz \g}}

\def\AT#1     {\ensuremath{A_{\mathrm{T}}^{#1}}\xspace}           
\def\btosgam  {\decay{\bquark}{\squark \g}}

\def\C#1      {\ensuremath{\mathcal{C}_{#1}}\xspace}                       
\def\Cp#1     {\ensuremath{\mathcal{C}_{#1}^{'}}\xspace}                    
\def\Ceff#1   {\ensuremath{\mathcal{C}_{#1}^{\mathrm{(eff)}}}\xspace}        
\def\Cpeff#1  {\ensuremath{\mathcal{C}_{#1}^{'\mathrm{(eff)}}}\xspace}       
\def\Ope#1    {\ensuremath{\mathcal{O}_{#1}}\xspace}                       
\def\Opep#1   {\ensuremath{\mathcal{O}_{#1}^{'}}\xspace}                    

\newcommand{\nospaceunit}[1]{\ensuremath{\text{#1}}}       
\newcommand{\aunit}[1]{\ensuremath{\text{\,#1}}}       

\newcommand{\tev}{\aunit{Te\kern -0.1em V}\xspace}
\newcommand{\gev}{\aunit{Ge\kern -0.1em V}\xspace}
\newcommand{\mev}{\aunit{Me\kern -0.1em V}\xspace}
\newcommand{\kev}{\aunit{ke\kern -0.1em V}\xspace}
\newcommand{\ev}{\aunit{e\kern -0.1em V}\xspace}
 
\newcommand{\mevc}{\ensuremath{\aunit{Me\kern -0.1em V\!/}c}\xspace}
\newcommand{\gevc}{\ensuremath{\aunit{Ge\kern -0.1em V\!/}c}\xspace}
\newcommand{\mevcc}{\ensuremath{\aunit{Me\kern -0.1em V\!/}c^2}\xspace}
\newcommand{\gevcc}{\ensuremath{\aunit{Ge\kern -0.1em V\!/}c^2}\xspace}
\newcommand{\gevgevcc}{\ensuremath{\gev^2\!/c^2}\xspace} 

\def\mm   {\aunit{mm}\xspace}

\def\mum  {\ensuremath{\,\upmu\nospaceunit{m}}\xspace}

\def\fb   {\ensuremath{\aunit{fb}}\xspace}
\def\invfb   {\ensuremath{\fb^{-1}}\xspace}

\def\ps   {\ensuremath{\aunit{ps}}\xspace}

\def\order{{\ensuremath{\mathcal{O}}}\xspace}

\def\gsim{{~\raise.15em\hbox{$>$}\kern-.85em
          \lower.35em\hbox{$\sim$}~}\xspace}
\def\lsim{{~\raise.15em\hbox{$<$}\kern-.85em
          \lower.35em\hbox{$\sim$}~}\xspace}

\def\sqs   {\ensuremath{\protect\sqrt{s}}\xspace}

\def\pt         {\ensuremath{p_{\mathrm{T}}}\xspace}

\def\et         {\ensuremath{E_{\mathrm{T}}}\xspace}

\def\rad{\aunit{rad}\xspace}

\def\evtgen     {\mbox{\textsc{EvtGen}}\xspace}

\def\geant      {\mbox{\textsc{Geant4}}\xspace}

\def\photos     {\mbox{\textsc{Photos}}\xspace}

\def\pythia     {\mbox{\textsc{Pythia}}\xspace}

\def\tell1  {TELL1\xspace}
\def\ukl1   {UKL1\xspace}

\usepackage{cite} 
\usepackage{mciteplus}

\usepackage{longtable} 

\begin{document}

\renewcommand{\thefootnote}{\fnsymbol{footnote}}
\setcounter{footnote}{1}


\begin{titlepage}
\pagenumbering{roman}

\vspace*{-1.5cm}
\centerline{\large EUROPEAN ORGANIZATION FOR NUCLEAR RESEARCH (CERN)}
\vspace*{1.5cm}
\noindent
\begin{tabular*}{\linewidth}{lc@{\extracolsep{\fill}}r@{\extracolsep{0pt}}}
\ifthenelse{\boolean{pdflatex}}
{\vspace*{-1.5cm}\mbox{\!\!\!\includegraphics[width=.14\textwidth]{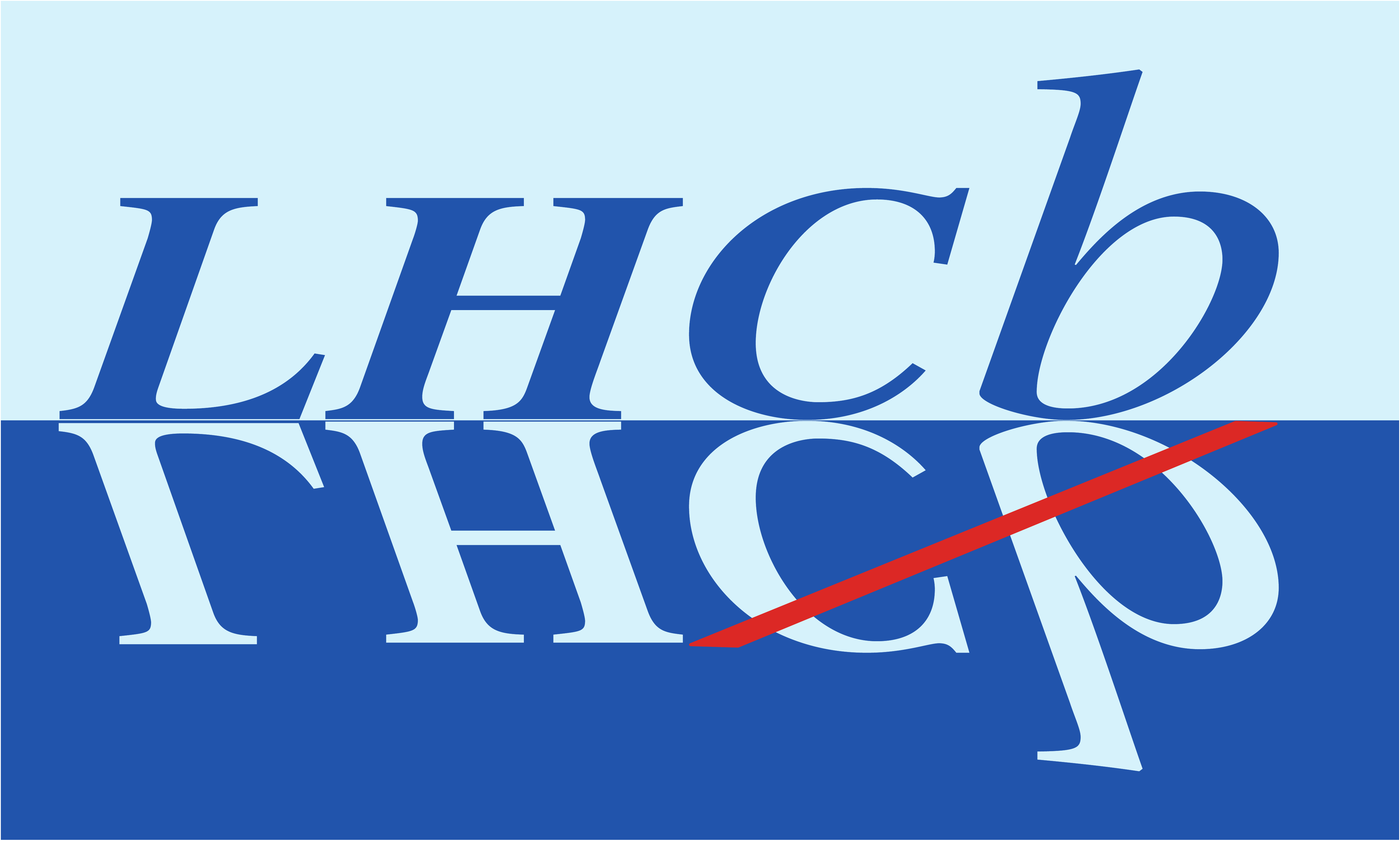}} & &}%
{\vspace*{-1.2cm}\mbox{\!\!\!\includegraphics[width=.12\textwidth]{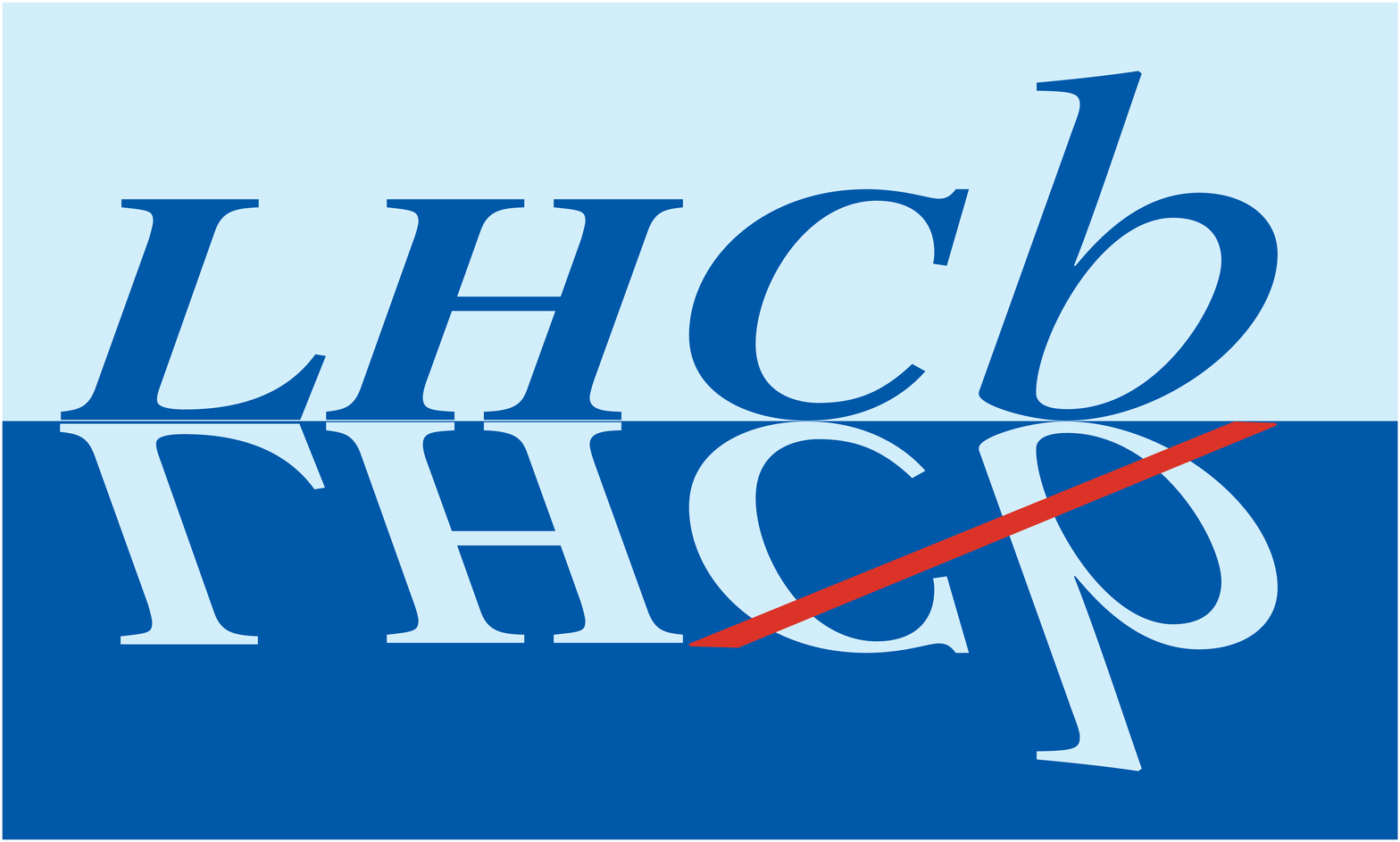}} & &}%
\\
 & & CERN-EP-2021-128 \\  
 & & LHCb-PAPER-2021-017 \\  
 & & January 17, 2022 \\ 
 & & \\
\end{tabular*}

\vspace*{4.0cm}

{\normalfont\bfseries\boldmath\huge
\begin{center}
  \papertitle 
\end{center}
}

\vspace*{2.0cm}

\begin{center}
\paperauthors\footnote{Authors are listed at the end of this paper.}
\end{center}

\vspace{\fill}

\begin{abstract}
  \noindent
  The first search for the rare radiative decay \XbGDec is performed using data collected by the \lhcb experiment in proton-proton collisions at a center-of-mass energy of 13\tev, corresponding to an integrated luminosity of 5.4\invfb. The \XbJDec channel is used as normalization. No \XbGDec signal is found and an upper limit of $\BR(\XbGDec) < 1.3 \times 10^{-4}$ at 95\% confidence level is obtained.
\end{abstract}

\vspace*{2.0cm}

\begin{center}
  Published in JHEP 01 (2022) 069
\end{center}

\vspace{\fill}

{\footnotesize 
\centerline{\copyright~\papercopyright. \href{\paperlicenceurl}{\paperlicence}.}}
\vspace*{2mm}

\end{titlepage}


\newpage
\setcounter{page}{2}
\mbox{~}
%
%
%
%


\renewcommand{\thefootnote}{\arabic{footnote}}
\setcounter{footnote}{0}

\cleardoublepage


\pagestyle{plain} 
\setcounter{page}{1}
\pagenumbering{arabic}



\section{Introduction}\label{sec:Intro}

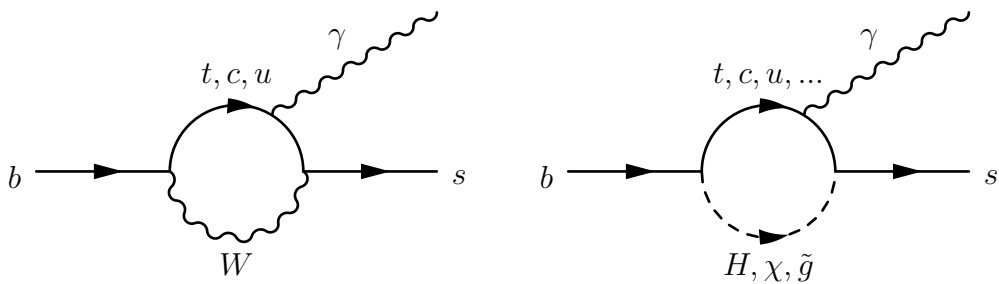
\begin{figure}[b]
    \centering
    \begin{fmffile}{penguin1}
        \begin{fmfgraph*}(150,75)  
            \fmftop{t1,t2}
            \fmfbottom{b,s}
            \fmf{fermion}{b,w1}
            \fmf{fermion,left,tension=0.5,label=$t,,c,,u$}{w1,w2}
            \fmf{fermion}{w2,s}
            \fmf{photon,right,tension=0.5,label=$W$}{w1,w2}
            \fmf{photon,label=$\gamma$}{ph,t2}
            \fmfforce{(0.59w,0.39h)}{ph}
            \fmflabel{$b$}{b}
            \fmflabel{$s$}{s}
          \end{fmfgraph*}
          \hspace{8ex}
        \begin{fmfgraph*}(150,75)  
            \fmftop{t1,t2}
            \fmfbottom{b,s}
            \fmf{fermion}{b,w1}
            \fmf{fermion,left,tension=0.5,label=$t,,c,,u,,...$}{w1,w2}
            \fmf{fermion}{w2,s}
            \fmf{scalar,right,tension=0.5,label=$H,,\chi,,\tilde{g}$}{w1,w2}
            \fmf{photon,label=$\gamma$}{ph,t2}
            \fmfforce{(0.59w,0.39h)}{ph}
            \fmflabel{$b$}{b}
            \fmflabel{$s$}{s}
          \end{fmfgraph*}
        \vspace{5ex}
    \end{fmffile}
    \vspace{1ex}
    \caption{\label{fig:penguin}
    The $b \to s\gamma$ penguin diagram, mediated by SM particles (left) and BSM particles (right).
    }
\end{figure}

The $\btosgam$ transition, depicted in Figure~\ref{fig:penguin}, is a flavor-changing neutral-current process characterized by the emission of a photon in the final state. Decays involving this feature are also known as radiative decays. 
The effective Hamiltonian in the operator product expansion formalism describing $\btosgam$ transitions at leading order is given by
\begin{equation}
    \mathcal{H}_{eff} = - \frac{G_F}{\sqrt{2}} V^*_{ts}V_{tb} (\mathcal{C}_7 \mathcal{O}_7 + \mathcal{C}_7' \mathcal{O}_7')\, ,
\end{equation}
where $\mathcal{O}_7$ $(\mathcal{O}_7')$ represents the left (right) projection of the electromagnetic penguin operator, which corresponds to the emission of a left (right)-handed photon. The strength of each contribution is encoded in the Wilson coefficients $\mathcal{C}_7$  and $\mathcal{C}_7'$. In the SM, the \Wm boson only couples to left-handed quarks and, thus, the only source of right-handed photons is due to helicity flips. 
The ratio of right- and left-handed amplitudes is expected to be $\mathcal{O}(m_s/m_b)$.
Thus, the SM predicts a negligible contribution of the right-handed operator $\mathcal{O}_7'$.
Measuring branching fractions, angular and charge-parity-violating observables in $\btosgam$ transitions enables testing the presence of right-handed contributions. Several analyses focusing on \B-meson decays have explored this field~\cite{LHCb-PAPER-2014-001, bFactories_B, LHCb-PAPER-2016-034, LHCb-PAPER-2019-015, LHCb-PAPER-2020-020}.

Radiative decays of \bquark-baryons provide access to the photon polarization due to the \mbox{spin \sfrac{1}{2}} ground state, the absence of flavor mixing and the presence of two spectator quarks. Therefore, \bquark-baryon decays provide complementary measurements to those performed with radiative \B-meson decays~\cite{bBaryons}.

The branching fraction of the \LbGDec decay mode has been recently measured for the first time~\cite{LHCb-PAPER-2019-010} and  constitutes the first radiative \bquark-baryon decay observed.\footnote{Charge-conjugated processes are implied throughout this paper.} Further radiative $b$-baryon decays can be studied with the \lhcb detector, providing complementary tests of the photon polarization in the SM. This paper focuses on the search for the \XbGDec decay mode, which is also mediated by the \btosgam transition. The \lhcb experiment provides unique conditions for studying the \XbGDec mode thanks to the large production of \bquark-baryons at the \lhc~\cite{LHCb-PAPER-2018-050, LHCb-PAPER-2018-047} and the excellent performance of the detector optimized for the analysis of \bquark-hadron decays. Additionally, previous measurements at \lhcb involving radiative B meson~\cite{LHCb-PAPER-2014-001, bFactories_B, LHCb-PAPER-2016-034, LHCb-PAPER-2019-015, LHCb-PAPER-2020-020} and \bquark-baryon~\cite{LHCb-PAPER-2019-010} decays motivates the search for new radiative baryonic modes, such as the \XbGDec decay.

The rare radiative $b$-baryon decay \XbGDec has not yet been observed. Using light-cone sum rules, its branching fraction, \mbox{\BR(\XbGDec)}, is predicted to be \mbox{$(3.03\pm0.10)\times10^{-4}$}~\cite{BR_pred}. This prediction is larger than the branching fraction of other radiative decays ($\BR\sim\order(10^{-5})$)~\cite{Bkpig, LHCb-PAPER-2012-019, LHCb-PAPER-2019-010}. A more recent study uses SU(3) flavor symmetry rules to predict $\BR(\XbGDec)=(1.23 \pm 0.64)\times10^{-5}$~\cite{New_BR_pred}. This second prediction uses the measurement of $\BR(\LbGDec)$ and thus it has a smaller dependency on estimated form factors. A measurement of the branching fraction of this decay could discriminate different approaches used in the theoretical predictions. This could help to estimate form-factors at low \qsq (photon pole) for the semileptonic decay \decay{\Xibm}{\Xim\mup\mun}~\cite{XibMuMu}. 
Furthermore, the possible signal obtained could be used to perform a measurement of the photon polarization~\cite{AngDists}. 

The data sample analyzed in this paper corresponds to an integrated luminosity of 5.4\invfb of proton-proton (\proton\proton) collisions at a center-of-mass energy of 13\tev, collected by the \lhcb detector. Potential experimenters' bias is avoided by validating the analysis procedure before inspecting the results. 
A normalization channel sharing the same hadronic part of the final state as the radiative decay is used to cancel potential systematic effects arising from detector efficiencies and the limited knowledge on the \Xibm production, $f_\Xibm$. The normalization channel is chosen to be the \XbJDec decay.

\section{LHCb detector}
The \lhcb detector~\cite{LHCb-DP-2008-001,LHCb-DP-2014-002} is a single-arm forward spectrometer covering the pseudorapidity range $2<\eta<5$, designed for the study of particles containing \bquark or \cquark quarks. The detector includes a high-precision tracking system consisting of a silicon-strip vertex detector surrounding the \proton\proton interaction region, a large-area silicon-strip detector located upstream of a dipole magnet with a bending power of about 4$\,$Tm, and three stations of silicon-strip detectors and straw drift tubes placed downstream of the magnet. The tracking system provides measurements of the momentum of charged particles with a relative uncertainty that varies from 0.5\% at low momentum to 1.0\% at 200\gevc. The minimum distance of a track to a primary \proton\proton collision vertex (PV), the impact parameter, is measured with a resolution of $(15\pm29/\pt)$\mum, where \pt is the component of the momentum transverse to the beam, in \gevc. The PV is reconstructed by forming a common vertex from a large number of tracks, consistent with originating from a pp collision~\cite{LHCb-PUB-2014-044}. Different types of charged hadrons are distinguished using information from two ring-imaging Cherenkov detectors. Photons, electrons and hadrons are identified by a calorimeter system consisting of scintillating-pad and preshower detectors, an electromagnetic and a hadronic calorimeter. Charged and neutral clusters in the electromagnetic calorimeter are discerned by extrapolating the tracks reconstructed by the tracking system to the calorimeter plane. Photons and neutral pions are distinguished by cluster shape, energy and mass distributions. Muons are identified by a system composed of alternating layers of iron and multiwire proportional chambers. Due to the photon energy resolution, \bquark-hadron decays with a high-energy photon in their final state are reconstructed with a \bquark-hadron mass resolution around 100\mevcc~\cite{LHCb-PAPER-2016-034}.

The online event selection is performed by a trigger system~\cite{LHCb-DP-2019-001}, consisting of a hardware stage, which uses information from the calorimeter and muon systems, followed by two software stages, which apply a partial and a full event reconstruction. At the hardware trigger stage, events are required to have a high-\pt photon or electron, detected in the electromagnetic calorimeter as a cluster of transverse energy (\et) with a threshold that varied between 2.1 and $3.0\gev$ during the data-taking period. The first stage of the software trigger requires a track well separated from any PV, and with a \pt higher than 1\gevc. At the second stage of the software trigger, the full decay chain is reconstructed to identify decays consistent with the signal mode. Only events signal candidate fulfilling the trigger requirements are kept.

Simulation is used to develop the selection strategy, compute the efficiency and determine the shape of the invariant-mass distribution of the signal decays. In the simulation, \proton\proton collisions are generated using \pythia~\cite{Sjostrand:2007gs,*Sjostrand:2006za} with a specific \lhcb configuration~\cite{LHCb-PROC-2010-056}. Decays of unstable particles are described by \evtgen~\cite{Lange:2001uf}, in which final-state radiation is generated using \photos~\cite{davidson2015photos}. The interaction of the generated particles with the detector, and its response are simulated using the \geant toolkit \cite{Allison:2006ve, *Agostinelli:2002hh}, as described in Ref.~\cite{LHCb-PROC-2011-006}. To save computing resources, the simulated signal decay is superimposed
to a limited set of simulated underlying interactions which are used multiple times~\cite{LHCb-DP-2018-004}.

\section{Selection}
 
\begin{figure}[tb]
    \centering
    \includegraphics[width=0.6\textwidth]{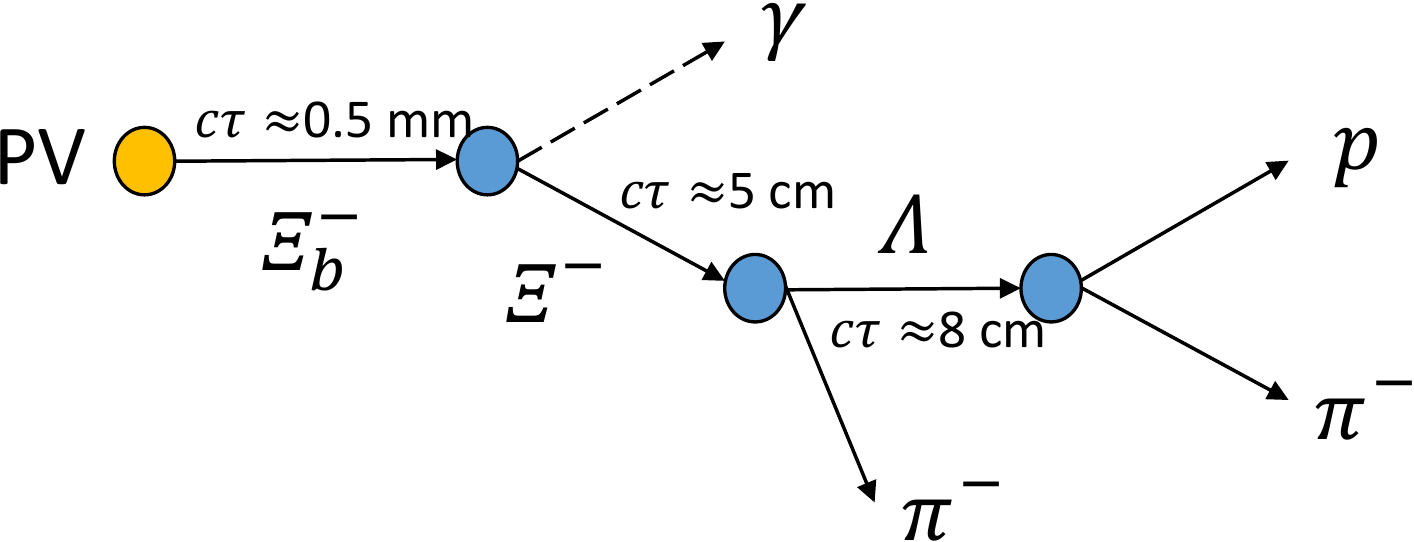}
    \caption{Topology of the \XbGDec decay, including three displaced vertices, with $c\tau$ of each particle given.}
    \label{fig:XibSketch}
\end{figure}

The reconstruction of the \XbGDec decay, with $\XDec$ and $\LDec$, involves the combination of two tracks with opposite charges originating from a common displaced vertex, and compatible with the $p$ and $\pi^-$ hypotheses. This is identified as a $\Lz$ baryon, which is combined with a $\pi^-$ track to form the $\Xim$ candidate. The $\Xibm$ candidate is in turn reconstructed as the combination of an energetic photon and the reconstructed $\Xim$ candidate. A sketch of the full decay chain, which includes three independent displaced vertices, is shown in Fig.~\ref{fig:XibSketch}.

\begin{figure}[tb]
    \centering
    \includegraphics[width=\CFigSizeI]{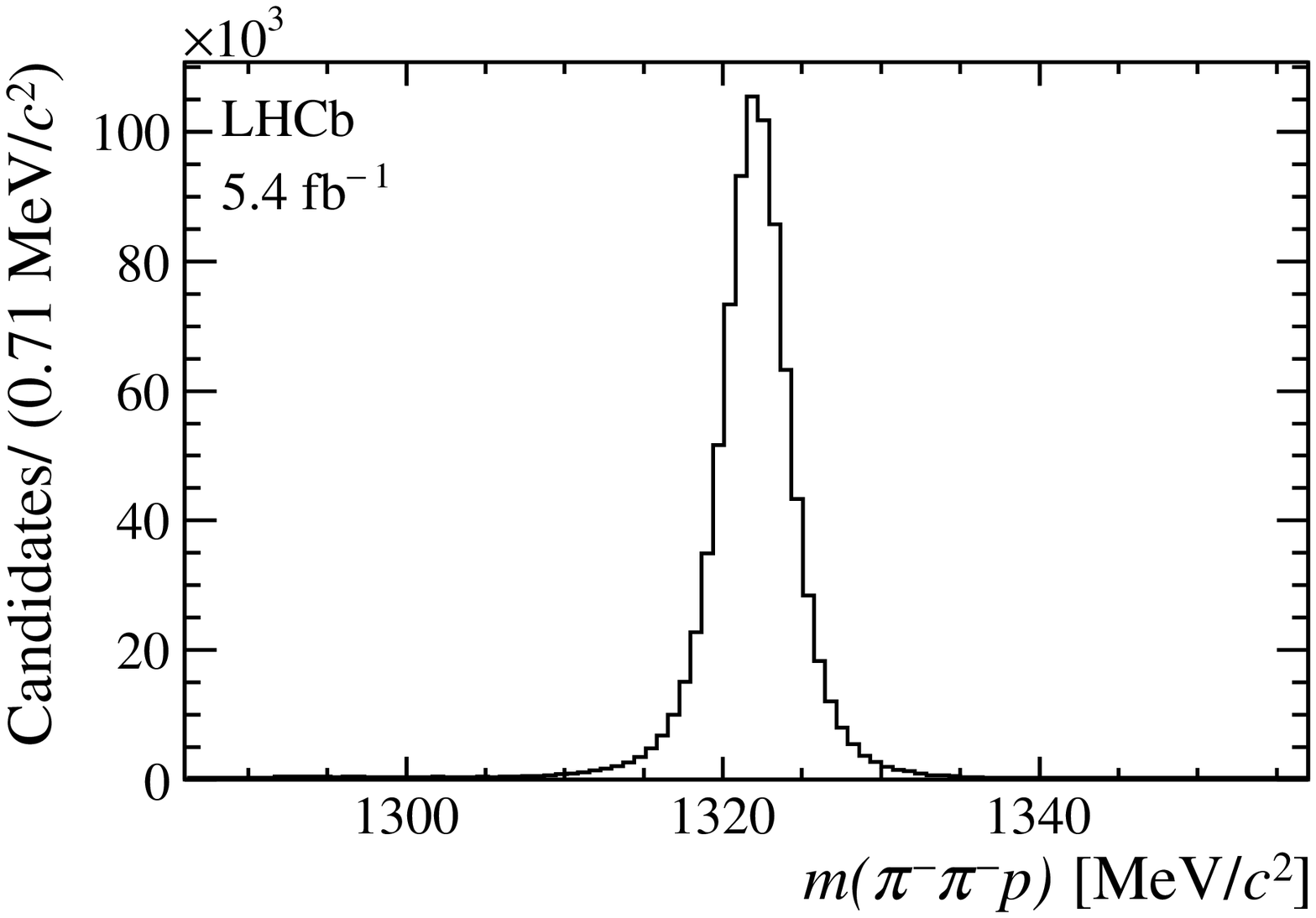}
    \caption{Mass distribution $m(\pim \pim\proton)$ showing the \decay{\Xim}{\Lz\pim} signal for events satisfying the trigger and the offline requirements described in the text for \XbGDec decays.}
    \label{fig:Xi_M}
\end{figure}

High-quality tracks inconsistent with originating from the PV are used for the reconstruction. For events with multiple PVs, the PV with the lowest impact parameter with respect to the candidate is used. Because of the long lifetime and large Lorentz boost, most of the \Lz and \Xim baryons decay outside the vertex detector. However, due to trigger limitations, only decays that occur inside the vertex detector can be considered. Particle identification requirements, based on a multivariate analysis technique, are applied to the charged particles~\cite{LHCb-DP-2018-001}. Proton and pion candidates with a minimum transverse momentum of 630\mevc and 250\mevc, respectively, are used to form a \Lz candidate. The proton-pion system is required to have a mass within 6\mevcc of the known \Lz mass~\cite{PDG2020} and a \pt larger than 1.5\gevc. The \Lz candidate is then combined with a pion candidate with $\pt > 130\mevc$ to form a \Xim candidate. This candidate is required to have decayed within 400\mm of the PV, to have a \pt larger than 2\gevc and a mass, $m(\pim\pim\proton)$ in the range \mbox{1310--1332\mevcc} around the known value of the \Xim mass of $1321.71\pm0.07\mevcc$~\cite{PDG2020}. After the trigger and offline requirements, a clean sample of \Xim candidates is obtained. The distribution of the mass of \Xim candidates is shown in Fig.~\ref{fig:Xi_M}.
Photon candidates are reconstructed from energy deposits in the electromagnetic
calorimeter not associated to any track. Background due to photons from \piz decays is rejected by a dedicated algorithm~\cite{LHCb-PUB-2015-016}. 
The \Xim candidate is combined with a photon candidate with \et larger than 4\gev. Due to the unknown photon direction and the long lifetime of the \Xim baryon, the \Xibm decay vertex cannot be determined. Consequently, the \Xibm trajectory is calculated assuming the photon  originates from the PV with the smallest distance of closest approach with respect to the \Xim trajectory. 
This is a good approximation given the short decay time of the \Xibm baryon. The \g  and \Xim momenta are then combined to reconstruct the \Xibm candidate. 
The \Xibm candidate must have \pt larger than 4\gevc and a mass within 800\mevcc of its known mass~\cite{PDG2020}. The distance of closest approach between the \Xibm and the \Xim trajectories must be $<50\mum$.

The combinatorial background, formed by random combinations of final-state particles, is suppressed by using a boosted decision tree (BDT)~\cite{Breiman}, employing the XGBOOST algorithm~\cite{xgboost}. The BDT classifier is trained using simulated samples of \XbGDec decays as signal, and candidates from data samples with the \Xim\g mass above 6.1\gevcc as a background proxy. The $k$-folding cross-validation technique~\cite{kFolding} with $k$ = 5 is used to avoid overfitting the BDT model. The variables used to train the BDT classifier are: the transverse momentum and the separation from the PV of the signal decay products; the photon pseudorapidity; the distance of closest approach between the \Xim decay products and between the \Xibm and \Xim flight directions; and the \pt asymmetry of the \Xim and the \g candidates. The \pt asymmetry for a given particle is computed as the normalized difference between the summed momenta of all  tracks within a cone of 1\rad around the particle direction, and the momentum of the particle.  The above variable discriminates against partially reconstructed backgrounds, consisting of decays with additional particles in the final state that have not been reconstructed. As the BDT classifier is trained using simulation, good agreement between the simulation and data is needed. This is validated using the \decay{\Lb}{\jpsi\proton\Km} and \decay{\Bz}{\Kstar\g} control modes employing the selection criteria described in Refs.~\cite{LHCb-PAPER-2016-049} and \cite{LHCb-PAPER-2016-034}, respectively. The normalization channel \XbJDec, with the selection described below, is also used for the same purpose. 
The event multiplicity, defined as the number of tracks per event, along with the \bquark-baryon momentum and transverse momentum are corrected for discrepancies between simulation and data. These corrections are extracted from \decay{\Lb}{\jpsi\proton\Km} background-subtracted data and simulated samples.
The BDT classifier is optimized by maximizing the Punzi figure of merit~\cite{Punzi:2003bu}, $\epsilon_\text{s}/(\sqrt{B} + 2.5)$, where $\epsilon_\text{s}$ is the efficiency of the requirement on the BDT output extracted from simulated signal events, and $B$ is the background yield from the high-mass sideband, extrapolated to the signal region. The chosen working point keeps 69\% of the signal candidates, while suppressing about 98\% of the combinatorial background.

The online reconstruction of candidates from the \XbJDec normalization channel, with $\jpsi \to \mup \mun$, follows a different strategy as compared to the \XbGDec signal channel. The muons, originating from inside the vertex detector, must pass the trigger for the normalization channel. This allows \Lz and \Xim baryons decaying both inside and outside the vertex detector to be used. 
For the trigger selection of the normalization channel, events are required to either have a muon with a \pt above 1.5\gevc, or two muons with a transverse momentum product greater than 1.6\gevgevcc. In the first software stage of the trigger, the event must have either a system of two well-identified oppositely charged muons with a large mass, $m(\mup \mun)>2.7\gevcc$, or at least one muon with $\pt>1\gev$ that is inconsistent with originating from any PV. In the second stage, events containing a \mup\mun-pair with a mass consistent with the known \jpsi mass~\cite{PDG2020}, and with a vertex significantly displaced from any PV, are selected.
The offline reconstruction follows similar criteria to the \XbGDec selection. The \jpsi candidate is reconstructed from two oppositely-charged tracks compatible with the muon hypothesis. The mass of the \mup\mun pair is required to be within a window of 60\mevcc around the known \jpsi mass~\cite{PDG2020}. In this case,  the \Xibm vertex is reconstructed with an improved \bquark-baryon mass resolution, with respect to the radiative decay, due to a precise measurement of the muon momenta. The \Xibm candidate is required to have a measured decay time between 0.3 and 1.4\ps, a mass within 300\mevcc of the \Xibm measured mass~\cite{PDG2020} and a good quality decay vertex. Given the high purity of the \XbJDec sample after the described selection, no BDT selection is used.

\section{Yield determination}

The signal is isolated from the background components by a fit to the reconstructed \Xibm mass distribution of the selected candidates. An unbinned maximum likelihood fit to the radiative, \XbGDec, and the normalization, \XbJDec, decay modes are used. The signal-mass shape is modeled with a double-sided Crystal Ball probability density function~\cite{Skwarnicki:1986xj}, comprising a Gaussian core and a power-law tail at both sides. The parameters for the tails are extracted from a fit to simulated samples. In the mass fit to data, the peak position for the radiative and normalization channels is the same, while the peak width is related using a scaling factor defined as the ratio of the signal and normalization widths in simulation. Sources of non-combinatorial background are investigated using simulated samples. The narrow width of the \Lz and \Xim baryons~\cite{PDG2020} and the clean sample of the latter (see Fig.~\ref{fig:Xi_M}) reduces the contamination from decays where one or more final state particles are misidentified, such as \mbox{\decay{\Omegab}{\Omegam\g}}. No candidates from the partially reconstructed background \mbox{\decay{\Xibm}{\Xi\eta}} with \mbox{\decay{\eta}{\g\g}} are expected in the selected data sample. 
There are no predictions for \Xibm baryons decaying into \piz mesons. This class of contamination is known to be suppressed in \Bz decays to \Kstar\g and \Kstar\piz final states, and the same is assumed in the baryon sector~\cite{PDG2020}. The only relevant background component is the combinatorial one, which is modeled with an exponential function. The mass fit is validated using pseudoexperiments with $\BR(\XbGDec)$ hypotheses ranging from $10^{-5}$ to $10^{-3}$.

The branching fraction is determined as
	\begin{align}\label{eq:Xb_BR}
		\BR(\XbGDec) & =  \BR(\XbJDec) \BR(J/\psi \to \mu^+ \mu^-) \nonumber \\ &   
		\times    
		\frac{\epsilon(\XbJDec)}{\epsilon(\XbGDec)} 
		\frac{N(\XbGDec)}{N(\XbJDec)}   \\ &
		= \alpha N(\XbGDec)\, , \nonumber
	\end{align}
where \BR indicates a branching fraction, $N$ is the signal yield extracted from the mass fit, $\epsilon$ denotes the combined reconstruction and selection efficiency for the given decay and $\alpha$ is the single-event sensitivity. Calibration samples of \decay{\Lz}{\proton\pim}, \decay{\Dz}{\Km\pip}, \decay{\jpsi}{\mup\mun} and \decay{\Bz}{\Kstar\g} are used to calculate the efficiencies of the particle identification requirements~\cite{LHCb-DP-2018-001, LHCb-PUB-2015-016}. The remaining selection and reconstruction efficiencies are determined from simulated samples.

 The value of the \XbJDec branching fraction multiplied by the hadronization fraction of \Xibm baryons, $f_\Xibm$, is provided in Ref.~\cite{PDG2020}. Due to the lack of precision in the $f_\Xibm$ absolute value, the \XbJDec branching fraction is computed using the SU(3) relation \mbox{$\Gamma(\XbJDec)=(3/2 \pm 0.45)\times\Gamma(\LbJDec)$}~\cite{LbXiB_SU3_1} instead. The quoted uncertainty is typical for flavor SU(3) predictions.
 \begin{table}[tb]
 \centering

\caption{Input parameters used to compute the branching fraction $\BR(\XbJDec)$.}
\label{tab:BR_Inputs}
\begin{tabular}{lr}
\hline
Parameter      & Value                                                        \\ \hline
$\tau_{\Xib}$  & $1.57 \pm 0.04\ps$~\cite{PDG2020}                            \\
$\tau_{\Lb}$   & $1.47 \pm 0.01\ps$~\cite{PDG2020}                            \\
$\BR(\LbJDec)$ & $(3.36 \pm 1.11)\times 10^{-4}$~\cite{Lb2L0Jpsi_CDF,HFLAV18} \\ \hline
\end{tabular}
\end{table}
 Combining the values listed in Tab.~\ref{tab:BR_Inputs}, the computed value of $\BR(\XbJDec)$ is 
\begin{align}\label{eq:BR_Xb2XiJpsi_SU3}
    \BR(\XbJDec) = \Big(\frac{3}{2} \pm 0.45\Big)\frac{\tau_{\Xib}}{\tau_{\Lb}} \BR(\LbJDec)
        =  (5.4 \pm 2.4)\times 10^{-4} \, .
\end{align}

\section{Results}\label{sec:mass_fit_Result}
\begin{figure}[htb]
	    \centering
        \includegraphics[width=\CFigSizeI]{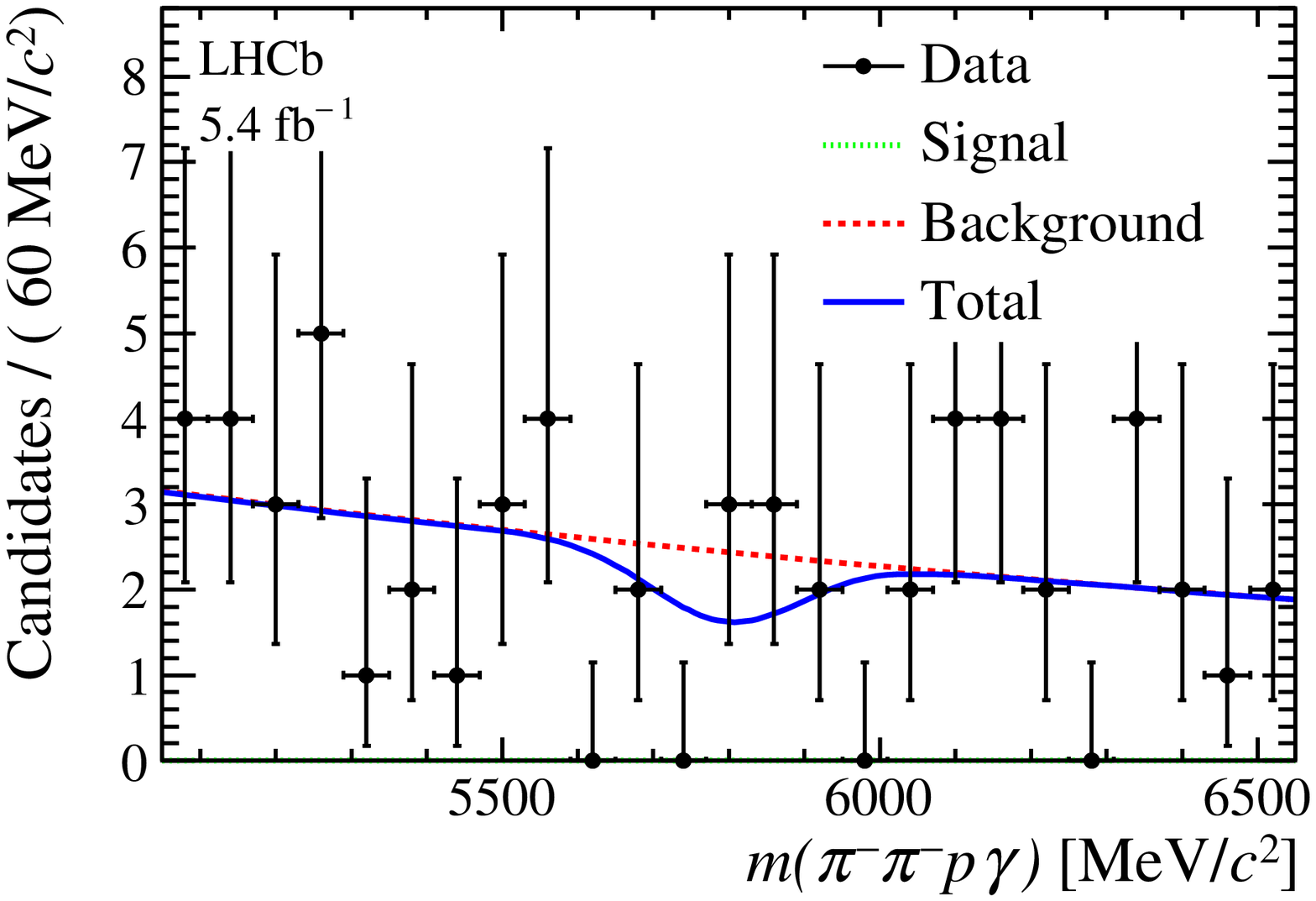}
        
        \includegraphics[width=\CFigSizeI]{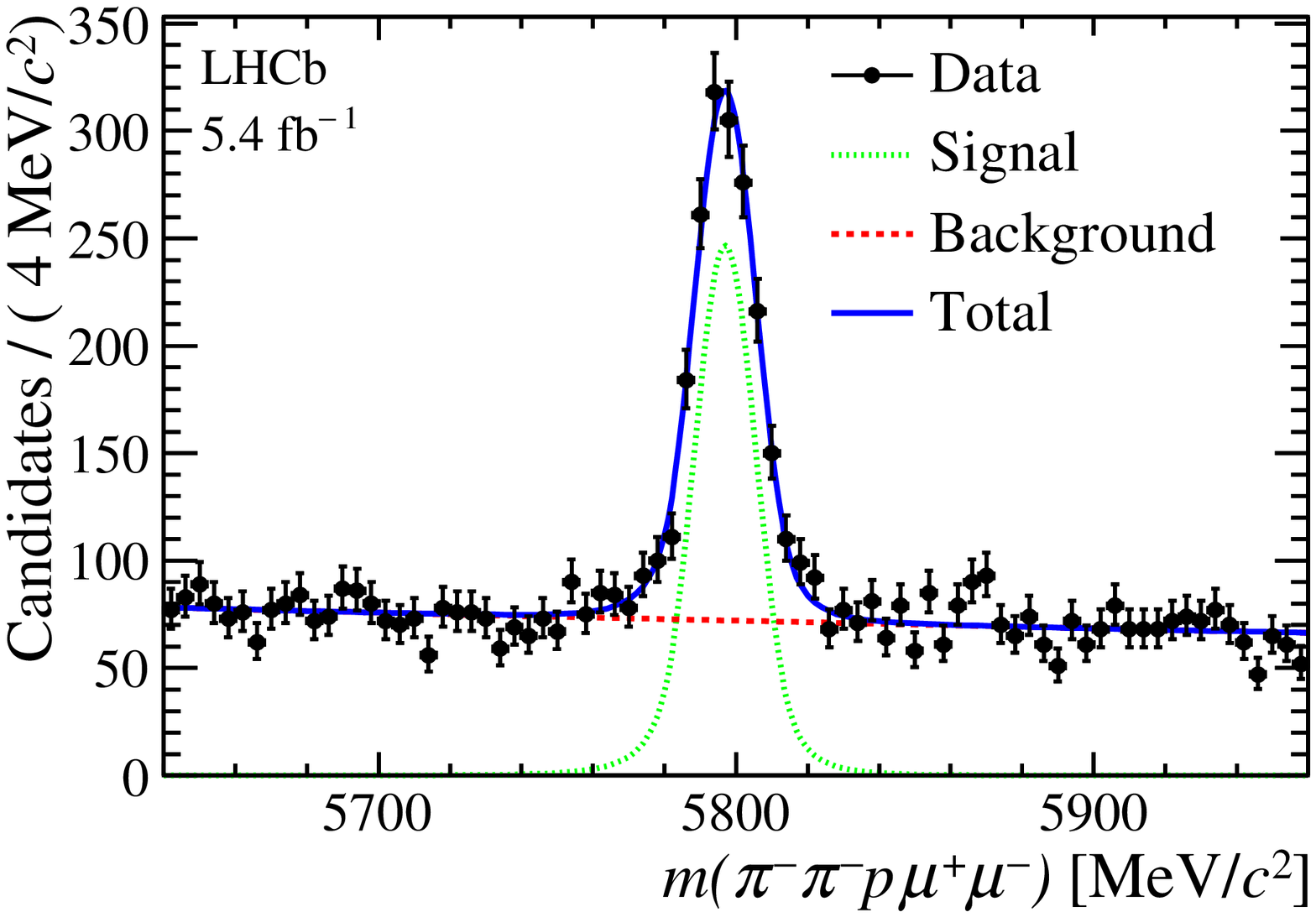}
        \caption{Distribution of (top) mass $m(\pim\pim\proton\gamma)$ for selected \Xim\g (\decay{\Xim}{\Lz\pim}) and (bottom) $m(\pim\pim\proton\mup\mun)$ for selected \Xim\jpsi (\decay{\jpsi}{\mup\mun}) candidates. The projections of the simultaneous fit are overlaid.}
	    \label{fig:Xb_massfit_RD}
	\end{figure}

Figure~\ref{fig:Xb_massfit_RD} shows the distribution of (top) the mass $m(\pim\pim\proton\gamma)$ for selected \Xim\g and (bottom) $m(\pim\pim\proton\,\mup\mun)$ for selected \Xim\jpsi candidates.
The simultaneous mass fit to these mass distributions returns  yields of \mbox{$N(\XbGDec) = -3.6\pm 3.9$} and \mbox{$N(\XbJDec) = 1407\pm52$}. Using the yield for the normalization channel together with the other quantities of Eq.~\ref{eq:Xb_BR}, a single-event sensitivity of $\alpha = (7.9\pm3.6)\times 10^{-6}$ is obtained.

Systematic uncertainties on the measurement of the $\BR(\XbGDec)$ value arise from several sources. The systematic effect due to the choice of the mass fit model is assessed by means of pseudoexperiments wherein the mass distribution is generated with an alternative model and fitted using the default model. The validation of fixing the value of the scale parameter relating the radiative and normalization channel resolutions is assessed by repeating the measurement considering possible differences between data and simulation. No deviation is found with respect to the nominal measurement and, thus, no systematic uncertainty is assigned to this effect. The uncertainty on the selection efficiencies, originating from the limited sample size, is propagated to the branching fraction and considered as a systematic uncertainty. The corrections applied to the simulation to improve the agreement with data are varied within their statistical uncertainty.
The effect of a possible mismodeling of the radiative hardware level trigger is assessed by comparing the efficiency extracted from simulation and from a method using the \BdKstGam decay as a control channel. The limited precision of the external value of $\BR(\XbJDec)$ induces the largest systematic uncertainty. Table~\ref{tab:syst} summarizes the systematic uncertainties.
\begin{table}[tb]
\centering
\caption{Dominant systematic uncertainties on the measurement of the branching fraction \BR(\XbGDec).}
\begin{tabular}{lr}
\hline 
Source            & Uncertainty (\%) \\ \hline
Mass fit model (signal)        & 9.1             \\
Mass fit model (background)    & 7.8             \\
Efficiency ratio               & 4.6              \\
Hardware trigger               & 10.0             \\
Simulation/Data agreement      & 6.0              \\
$\BR(\XbJDec)$                 & 45.6             \\ \hline 
Sum in quadrature              & 48.7             \\ \hline 
\end{tabular}
\label{tab:syst}
\end{table}

\begin{figure}[htb]
    \centering
    \includegraphics[width=\CFigSizeI]{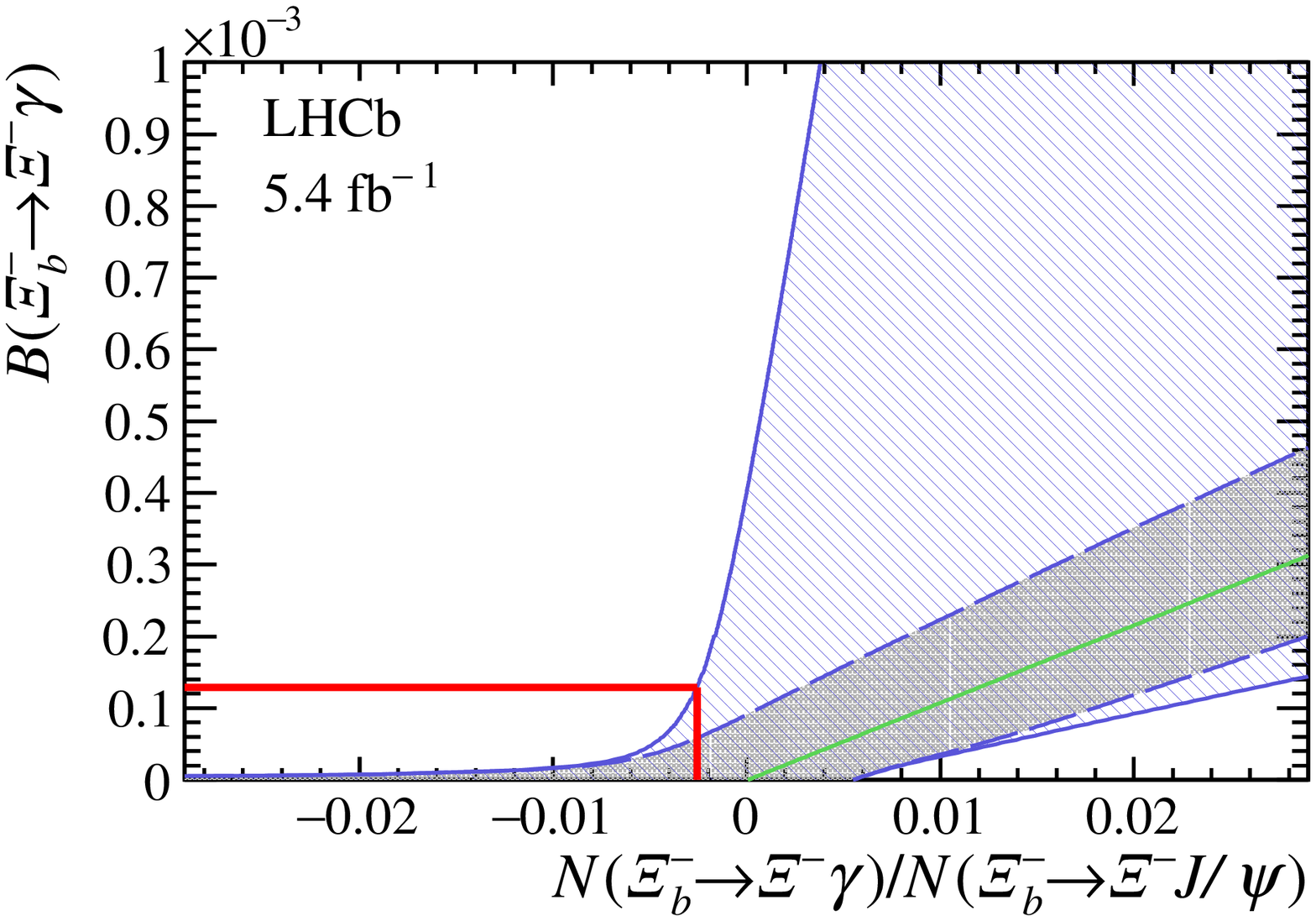}
    \caption{Confidence interval at 95\% CL showing the upper limit for $\BR(\XbGDec)$ as a function of the $\frac{N(\XbGDec)}{N(\XbJDec)}$ ratio. The green line represents the relation between the yield and the branching fractions. The interval considering only the statistical uncertainty is shown by the dashed blue lines, while the full blue lines also includes the systematic uncertainties. The measured ratio of the yields and the upper limit on the branching fraction are represented by the red line.}
    \label{fig:Xb_BR_FC}
\end{figure}

Since no \XbGDec signal is observed, the Feldman-Cousins (FC) method~\cite{FeldmannCousins} is used to set an upper limit on the value of $\BR(\XbGDec)$. 
For the FC method, the relation between the true and fitted signal yield and the statistical uncertainty are determined from pseudoexperiments. The systematic uncertainty is added in quadrature to the statistical uncertainty. The value of $\BR(\XbGDec)$ is calculated from the signal yield using Eq.~(\ref{eq:Xb_BR}). From this set of values, the 95\% confidence level (CL) is built and shown in Fig.~\ref{fig:Xb_BR_FC}.

Combining this study with the measured yield ratio, an upper limit is set
\begin{equation*}
     \BR(\XbGDec) < 1.3 \, (0.6) \times 10^{-4} \; \text{ at } 95\% \, (90\%) \text{ CL}\,  .
\end{equation*}

This is the first limit on this decay channel. 
Because the systematic uncertainty from the normalization channel branching fraction is dominant, the ratio of the branching fractions is reported, where the total systematic reduces to 17\%. Using the FC approach, an upper limit of

\begin{equation*}
     \frac{\BR(\XbGDec)}{\BR(\XbJDec)} < 0.12\, (0.08) \text{ at } 95\% \, (90\%) \text{ CL}\, , 
\end{equation*}
is set.

\section{Conclusion}

The first search for \bquark-baryon flavor-changing neutral-current  
radiative \XbGDec decay is reported, using \proton\proton collision data at a
center-of-mass energy of $\sqs=13\tev$ collected by the \lhcb experiment. The data set 
corresponds to an integrated luminosity of 5.4\invfb. No evidence for a signal is 
found. Upper limits at 90\% and 95\% CL of the value of $\BR(\XbGDec)$ are reported, which 
are in slight tension with the predictions from light-cone sum rules~\cite{BR_pred} but are consistent with flavor-symmetry 
driven predictions from Ref.~\cite{New_BR_pred}.


\section*{Acknowledgements}
%
%
\noindent We express our gratitude to our colleagues in the CERN
accelerator departments for the excellent performance of the LHC. We
thank the technical and administrative staff at the LHCb
institutes.
We acknowledge support from CERN and from the national agencies:
CAPES, CNPq, FAPERJ and FINEP (Brazil); 
MOST and NSFC (China); 
CNRS/IN2P3 (France); 
BMBF, DFG and MPG (Germany); 
INFN (Italy); 
NWO (Netherlands); 
MNiSW and NCN (Poland); 
MEN/IFA (Romania); 
MSHE (Russia); 
MICINN (Spain); 
SNSF and SER (Switzerland); 
NASU (Ukraine); 
STFC (United Kingdom); 
DOE NP and NSF (USA).
We acknowledge the computing resources that are provided by CERN, IN2P3
(France), KIT and DESY (Germany), INFN (Italy), SURF (Netherlands),
PIC (Spain), GridPP (United Kingdom), RRCKI and Yandex
LLC (Russia), CSCS (Switzerland), IFIN-HH (Romania), CBPF (Brazil),
PL-GRID (Poland) and NERSC (USA).
We are indebted to the communities behind the multiple open-source
software packages on which we depend.
Individual groups or members have received support from
ARC and ARDC (Australia);
AvH Foundation (Germany);
EPLANET, Marie Sk\l{}odowska-Curie Actions and ERC (European Union);
A*MIDEX, ANR, IPhU and Labex P2IO, and R\'{e}gion Auvergne-Rh\^{o}ne-Alpes (France);
Key Research Program of Frontier Sciences of CAS, CAS PIFI, CAS CCEPP, 
Fundamental Research Funds for the Central Universities, 
and Sci. \& Tech. Program of Guangzhou (China);
RFBR, RSF and Yandex LLC (Russia);
GVA, XuntaGal and GENCAT (Spain);
the Leverhulme Trust, the Royal Society
 and UKRI (United Kingdom).




\addcontentsline{toc}{section}{References}
\bibliographystyle{LHCb}
\bibliography{main,standard,LHCb-PAPER,LHCb-CONF,LHCb-DP,LHCb-TDR}

\newpage
\centerline
{\large\bf LHCb collaboration}
\begin
{flushleft}
\small
R.~Aaij$^{32}$,
A.S.W.~Abdelmotteleb$^{56}$,
C.~Abell{\'a}n~Beteta$^{50}$,
T.~Ackernley$^{60}$,
B.~Adeva$^{46}$,
M.~Adinolfi$^{54}$,
H.~Afsharnia$^{9}$,
C.A.~Aidala$^{86}$,
S.~Aiola$^{25}$,
Z.~Ajaltouni$^{9}$
% ...

{\footnotesize \it

% ...
$^{9}$Universit{\'e} Clermont Auvergne, CNRS/IN2P3, LPC, Clermont-Ferrand, France\\
% ...
$^{25}$INFN Sezione di Milano, Milano, Italy\\
% ...
$^{32}$Nikhef National Institute for Subatomic Physics, Amsterdam, Netherlands\\
% ...
$^{46}$Instituto Galego de F{\'\i}sica de Altas Enerx{\'\i}as (IGFAE), Universidade de Santiago de Compostela, Santiago de Compostela, Spain\\
% ...
$^{50}$Physik-Institut, Universit{\"a}t Z{\"u}rich, Z{\"u}rich, Switzerland\\
% ...
$^{54}$H.H. Wills Physics Laboratory, University of Bristol, Bristol, United Kingdom\\
% ...
$^{56}$Department of Physics, University of Warwick, Coventry, United Kingdom\\
% ...
$^{60}$Oliver Lodge Laboratory, University of Liverpool, Liverpool, United Kingdom\\
% ...
$^{68}$Syracuse University, Syracuse, NY, United States\\
% ...
$^{86}$University of Michigan, Ann Arbor, United States, associated to $^{68}$\\
\bigskip
% ...
\medskip
}
\end{flushleft}

\end{document}